\documentclass[11pt,letterpaper]{article} 

\pdfoutput=1
\usepackage{amsmath,amssymb,amsfonts}
\usepackage[usenames,dvipsnames]{xcolor}
\definecolor{airforceblue}{rgb}{0.36, 0.54, 0.66}
\definecolor{meb}{rgb}{0.01, 0.31, 0.59}
\usepackage[T1]{fontenc}
\usepackage{palatino}
\def\title{Radiation in Holography} 
\usepackage{graphicx}
\usepackage[colorlinks=true, pdfstartview=FitV, linkcolor=blue, citecolor=magenta, urlcolor=blue]{hyperref}
\usepackage{cancel}
\usepackage{mdframed,framed}
\usepackage[usenames,dvipsnames]{xcolor}
\usepackage{mathpazo}
\usepackage[normalem]{ulem}
\usepackage{layout}
\usepackage[nosort]{cite}

\newcommand{\beq}{\begin{eqnarray}}
\newcommand{\eeq}{\end{eqnarray}}
\newcommand{\beqn}{\begin{eqnarray}}
\newcommand{\eeqn}{\end{eqnarray}}
\newcommand{\pa}{\partial}

\newcommand{\cL}{{\cal L}}

\usepackage{tikz-cd}
\usepackage{pict2e}
\makeatletter
\newcommand{\variable@rule}[1]{%
  \fontdimen8  
  \ifx#1\displaystyle\textfont3\else
    \ifx#1\textstyle\textfont3\else
      \ifx#1\scriptstyle\scriptfont3\else
        \scriptscriptfont3\relax
  \fi\fi\fi
}
\usepackage{mathtools}

\usepackage{tocloft}
\newcommand{\ve}{\varepsilon}

\newcommand{\cP}{{\cal{{P}}}}

\newcommand{\bz}{{\overline{z}}}

\newcommand{\bzeta}{{\overline{\zeta}}}

\newcommand{\rd}{\text{d}}

\newcommand{\chkM}{{\color{red} \,\checkmark\kern-5pt{}_{M}}}

\newcommand{\be}{\begin{equation}}
\newcommand{\ee}{\end{equation}}
\newcommand{\bea}{\begin{eqnarray}}
\newcommand{\eea}{\end{eqnarray}}

\def\pa{\partial}

\def\bw{{\bar w}}

 \newcommand{\badat}{\begin{alignedat}}
 \newcommand{\eadat}{\end{alignedat}}
 
\def\p{\partial}

\usepackage[mathscr]{eucal}

\usepackage{color}

\newcommand{\pink}[1]{\textcolor{\pink}{#1}}

\definecolor{dblue}{rgb}{0.2,0.50,0.80}

\newcommand{\perimeter}[1]{
	\centerline{
		\begin{minipage}[c]{0.7\textwidth}
			\begin{center}
			$^a$ Perimeter Institute for Theoretical Physics,\\
			 31 Caroline St. N., Waterloo ON, Canada, N2L 2Y5
			\end{center}
		\end{minipage}
		}
	}
\newcommand{\berkeley}[1]{
	\centerline{
		\begin{minipage}[c]{0.8\textwidth}
			\begin{center}
		$^b$ Center for Theoretical Physics and Department of Physics, \\
        University of California, Berkeley, CA 94720, U.S.A.
			\end{center}
		\end{minipage}
		}
	}

\usepackage{amsmath,amssymb,amsfonts}
\usepackage[usenames,dvipsnames]{xcolor}
\definecolor{airforceblue}{rgb}{0.36, 0.54, 0.66}
\definecolor{meb}{rgb}{0.01, 0.31, 0.59}
\usepackage[T1]{fontenc}
\usepackage{palatino}
\usepackage{graphicx}
\usepackage[colorlinks=true, pdfstartview=FitV, linkcolor=blue, citecolor=magenta, urlcolor=blue]{hyperref}
\usepackage{cancel}
\usepackage{mdframed,framed}
\usepackage[usenames,dvipsnames]{xcolor}
\usepackage{mathpazo}
\usepackage[normalem]{ulem}
\usepackage{layout}
\usepackage[nosort]{cite}
\newcommand{\bg}{{\overline{g}}}
\newcommand{\bnabla}{{\overline{\nabla}}}

\newcommand{\sC}{{\mathscr{C}}}
\newcommand{\bR}{{\overline{R}}}
\newcommand{\bS}{{\overline{S}}}

\begin{document}

{\centering
 \vspace*{1cm}
\textbf{\LARGE{\title{}}}
\vspace{0.5cm}
\begin{center}
Luca Ciambelli,$^a$ Sabrina Pasterski,$^a$ Elisa Tabor$^{b}$\\
\vspace{0.5cm}
\textit{\perimeter{}}\\
\vspace{0.5cm}
\textit{\berkeley{}}
\end{center}
\vspace{0.5cm}
{\small{\href{mailto:ciambelli.luca@gmail.com}{ciambelli.luca@gmail.com}, \ \href{mailto:spasterski@perimeterinstitute.ca}{spasterski@perimeterinstitute.ca}, \  \href{mailto:etabor@berkeley.edu}{etabor@berkeley.edu}}}
\vspace{1cm}
\begin{abstract}
\vspace{0.5cm}
We show how to encode the radiative degrees of freedom in $4$-dimensional asymptotically AdS spacetimes, using the boundary Cotton and stress tensors. Background radiation leads to a reduction of the asymptotic symmetry group, in contrast to asymptotically flat spacetimes, where a non-vanishing news tensor does not restrict the asymptotic symmetries. Null gauges,  such as 
$\Lambda$-BMS,
provide a framework for AdS spacetimes that include radiation in the flat limit. We use this to check that the flat limit of the radiative data matches the expected definition in intrinsically asymptotically flat spacetimes. We further dimensionally reduce our construction to the celestial sphere, and show how the $2$-dimensional celestial currents can be extracted from the $3$-dimensional boundary data.
\end{abstract}}

\thispagestyle{empty}

\newpage
\tableofcontents
\thispagestyle{empty}
\newpage
\clearpage
\pagenumbering{arabic} 

\section{Introduction}

The holographic principle is a powerful tool for understanding the nature of quantum gravity. It posits that such theories should have an equivalent lower dimensional description in terms of a non-gravitational  quantum theory living on the conformal boundary.  While a major success of string theory is its ability to realize explicit top-down constructions of such dualities in the context of Anti-de Sitter (AdS) spacetimes\cite{Maldacena:1997re, Witten:1998qj}, this principle traces its roots to studies of black hole thermodynamics \cite{Bekenstein:1972tm,Hawking:1974rv} suggesting it should generalize to more astrophysically relevant spacetimes.

As reviewed in \cite{Strominger:2017zoo,Pasterski:2021raf}, the Celestial Holography program is a recent attempt to extend the holographic principle to asymptotically flat spacetimes, which describe the universe at a wide range of scales: from scattering experiments to gravitational wave observations. However, the flat limit, that is, the vanishing cosmological constant limit, is subtle. Not only do the compact extra dimensions get large in this limit -- spoiling top-down constructions -- even a bottom up approach runs into issues. Whereas the boundary of global  AdS is a timelike cylinder, the conformal compactification of Minkowski contains two null components: $\mathcal{I}^-$ and $\mathcal{I}^+$ where massless particles respectively enter and exit the spacetime. Moreover, while AdS is reminiscent of putting gravity in a box with reflective boundary conditions, this lack of leaking radiation complicates our ability to achieve a physical flat limit, and impacts the utility of AdS/CFT for studying black hole evaporation \cite{Penington:2019npb, Almheiri:2019psf}.

The goal of this paper is to understand how to describe radiation in AdS spacetimes and then connect their flat limit to the Celestial Hologram. Proposing an extension of the result of Fern\'andez-\'Alvarez and Senovilla \cite{Fernandez-Alvarez:2019kdd, Fernandez-Alvarez:2020hsv, Fernandez-Alvarez:2021yog, Fernandez-Alvarez:2021zmp, Fernandez-Alvarez:2021uvz, Senovilla:2022pym} to AdS, we show how asymptotic radiation is encoded in the super-Poynting vector whose boundary value, in four bulk dimensions, is proportional to the commutator between the boundary Cotton-York tensor and the boundary stress tensor. This implies that both must be non-zero in order to have radiation, which has important consequences for the symmetries of the AdS hologram. The standard AdS/CFT dictionary is formulated so that the bulk metric approaches a conformally flat structure at the conformal boundary, and thus a vanishing Cotton tensor. The duality is more complicated when the boundary metric is not conformally flat, see e.g. \cite{Martelli:2011fu, Parisini:2022wkb}, and indeed, it is still an open question how to include the Cotton tensor in the boundary CFT. It is part of our agenda to demonstrate in which precise sense the conditions leading to AdS/CFT relate to the background geometric radiation. In particular, we will see that the presence of radiation requires a non-vanishing Cotton tensor which implies that the number of boundary conformal isometries, that is, the number of asymptotic symmetries from the bulk with Dirichlet boundary conditions, is reduced. 

This reduction of the asymptotic symmetry group in the presence of radiation is in stark contrast to asymptotically flat spacetimes, where a non-vanishing news tensor does not restrict the asymptotic symmetries. In fact, the asymptotic symmetry group is enhanced from the naive flat limit leading 
to the BMS group~\cite{Bondi:1962px,Sachs:1962wk,Sachs:1962zza,Barnich:2011ct}, independently of the presence of radiation. To perform this flat limit, we cannot use the Fefferman-Graham gauge \cite{Fefferman:1985, Fefferman:2007rka}, which is singular. Rather, this has been performed in null gauges (in $3$ bulk dimensions in \cite{Barnich:2012aw, Campoleoni:2018ltl, Ruzziconi:2020wrb, Geiller:2021vpg} and in the fluid/gravity correspondence in \cite{Ciambelli:2018wre, Campoleoni:2023fug}), in particular in Bondi gauge in \cite{Poole:2018koa} and in the $\Lambda$-BMS construction of~\cite{Compere:2019bua,Compere:2020lrt}. The latter provides an explicit form of asymptotically AdS metrics that limit to radiative spacetimes in the $\Lambda\rightarrow0$ limit. For these spacetimes, the time dependence of the boundary metric is suppressed by a factor of the cosmological constant that we are taking to zero. Meanwhile the components of the boundary stress tensor are singular in this limit.
Given its direct relevance for Celestial Holography, we use the $\Lambda$-BMS setup to first examine the flat limit of our proposed super-Poynting vector. This naturally lands on a Carrollian description of flat space holography \cite{Ciambelli:2018wre, Donnay:2022aba,Donnay:2022wvx}, where the hologram has codimension-$1$ and the null time is part of the dual description. We then relate the celestial currents for both the BMS~\cite{He:2014laa,Kapec:2014opa,Kapec:2016jld} and $w_{1+\infty}$~\cite{Guevara:2021abz,Strominger:2021mtt} symmetries to the boundary Cotton and stress tensors by dimensionally reducing to the celestial sphere. We note that this is in the context of different boundary conditions for the AdS metric than in the recent related work~\cite{deGioia:2023cbd}. This sheds light on how the codimension-$2$ celestial hologram relates to its more standard codimension-$1$ AdS analog. 

This paper is organized as follows. In section~\ref{sec:ads} we propose a geometric definition for radiation in asymptotically AdS spacetimes. Specifically, we extend the construction of \cite{Fernandez-Alvarez:2019kdd, Fernandez-Alvarez:2020hsv, Fernandez-Alvarez:2021yog, Fernandez-Alvarez:2021zmp, Fernandez-Alvarez:2021uvz, Senovilla:2022pym} to AdS in section~\ref{sec:adsrad} and then examine the consequences of non-trivial radiation on the AdS/CFT correspondence in section~\ref{sec:adscft}.  We then turn to the flat hologram in section~\ref{sec:flat}, reviewing the asymptotic structure in section~\ref{sec:bondi}. We use the $\Lambda$-BMS construction of~\cite{Compere:2019bua,Compere:2020lrt} to examine the flat limit of our radiation proposal in section~\ref{sec:ads2flat}, and connect the AdS stress tensor to the celestial symmetry generators in section~\ref{sec:ccft}. Finally, we close with a discussion in section~\ref{sec:conclusions}.

\section{AdS}\label{sec:ads}

In this Section we describe a proposal for the notion of radiation in AdS spacetimes. In the first part, we discuss how the results of Fern\'andez-\'Alvarez and Senovilla \cite{Fernandez-Alvarez:2019kdd, Fernandez-Alvarez:2020hsv, Fernandez-Alvarez:2021yog, Fernandez-Alvarez:2021zmp, Fernandez-Alvarez:2021uvz, Senovilla:2022pym} can be exported to AdS, giving a geometric notion of radiation therein. In the second part, we review some basics of AdS/CFT, and we comment on how the usual holographic framework is based on gravitational non-radiative backgrounds, in order to exploit the full power of asymptotic symmetries. We conclude analysing how to introduce  background radiation and the consequent challenges. 

\subsection{Radiation in AdS}\label{sec:adsrad}

In the following, we consider $4$-dimensional bulk spacetimes with negative cosmological constant. We use indices $\{\mu,\nu,\dots\}$ for the bulk and $\{a,b,\dots\}$ for the $3$-dimensional boundary.

\paragraph{Asymptotic Structure}

A $4$-dimensional asymptotically AdS spacetime (aAdS$_4$) is a solution of the Einstein equations with negative cosmological constant
\beq\label{eq:eeads}
G_{\mu\nu}+\Lambda g_{\mu\nu}=8\pi G_N T^M_{\mu\nu}.
\eeq
The conformal boundary is a timelike hypersurface and the bulk metric has a second order pole approaching the boundary, as discussed by Penrose \cite{Penrose:1962ij, Penrose:1964ge} (see also \cite{Geroch:1977big}). 

A convenient gauge to describe aAdS$_4$ spaces is the Fefferman-Graham gauge \cite{Fefferman:1985, Fefferman:2007rka} (see also \cite{Henningson:1998gx, Skenderis:1999nb}). Calling $r$ the radial coordinate so that $x^\mu=(r,x^a)$, we have
\beq \label{FG}
\rd s^2=g_{ab}(r,x)\rd x^a \rd x^b+\frac{\rd r^2}{r^2}=\sum^{\infty}_{n=0}\left(r^{2-2n}g^{(2n)}_{ab}(x)+r^{-1-2n}\pi^{(2n)}_{ab}(x)\right)\rd x^a \rd x^b+\frac{\rd r^2}{r^2},
\eeq 
where we set the AdS radius to one.\footnote{The relationship between the cosmological constant and the AdS radius is $\ell_{AdS}^2=\frac{3}{|\Lambda|}$.} This gauge is achieved imposing the conditions $g_{rr}=\frac1{r^2}$ and $g_{ra}=0$. The conformal boundary ${\cal B}$ is located at $r\to \infty$ and the bulk induces a conformal class of boundary metrics $[g^{(0)}_{ab}(x)]$. From now on, the boundary metric is denoted by $\bg_{ab}=g^{(0)}_{ab}$, its Levi-Civita connection is $\bnabla$ and its curvature tensors have always an overline. 

The radial equations of motion give recursion relations for $g^{(2n)}$ and $\pi^{(2n)}$ such that the phase space is entirely determined by $\bg_{ab}$ and $\pi^{(0)}_{ab}$. In the absence of matter, the latter is further constrained to satisfy
\beq\label{pi}
\bnabla_a \pi^{(0)a}{}_b=0, \qquad Tr(\pi^{(0)})=0,
\eeq
where indices are raised using the metric $\bg_{ab}$. The tensors $\bg_{ab}$ and $\pi^{(0)}_{ab}$ are the first and rescaled, i.e., first non-vanishing term in the radial expansion, second fundamental forms of the (conformally compactified) boundary, and thus can be treated as the configuration variable and conjugate momentum for the radial Hamiltonian problem, as described by Fefferman and Graham \cite{Fefferman:1985, Fefferman:2007rka}, similar to the de Sitter and flat analysis of Friedrich \cite{Friedrich:1981wx, Friedrich:1981at, Friedrich:1986qfi, Friedrich:1995vb}. In particular, $\pi^{(0)}_{ab}$ is the Brown-York tensor of the conformally rescaled boundary metric \cite{Brown:1992br, Balasubramanian:1999re}. In view of its holographic interpretation, we refer to $\pi^{(0)}_{ab}$ as the boundary stress tensor.

The infinitesimal symmetries $\xi(r,x)$ preserving the gauge conditions are found solving $\cL_\xi g_{rr}=0=\cL_\xi g_{ra}$ and are given by\footnote{See \cite{Alessio:2020ioh, Fiorucci:2020xto} for a study of asymptotic symmetries and boundary conditions in FG gauge.} 
\beq\label{adsxi}
\xi^r(r,x)=\sigma(x) r,\qquad \xi^a(r,x)=Y^a(x)-\pa_b\sigma(x) \int \frac{\rd r'}{r'}g^{ab}(r,x).
\eeq
Without further conditions on the falloffs, the bulk gauge-preserving diffeomorphisms (also called Penrose-Brown-Henneaux transformations \cite{Imbimbo:1999bj}) are thus encoded in the radial rescaling $\sigma(x)$ and  boundary diffeomorphisms $Y^a(x)$. The radial rescaling induces a Weyl transformation of the conformal boundary, allowing us to span through the conformal class of boundary metrics $[\bg_{ab}(x)]$.\footnote{However, radial transformations are always entangled with subleading hypersurface diffeomorphisms in the FG gauge. These effects can be disentangled working in an extended gauge, named the Weyl-Fefferman-Graham gauge \cite{Ciambelli:2019bzz}, further studied in \cite{Jia:2021hgy, Jia:2023gmk, Ciambelli:2023ott}.}

\paragraph{Cotton Tensor} Given that the bulk induces a conformal class of boundary metrics, the conformal properties of the boundary are important in holography. For $d>3$, the conformal tensor of a $d$-dimensional manifold is the Weyl tensor. In $d=3$, which is the case for our boundary, the Weyl tensor identically vanishes and its role is played by the Cotton tensor \cite{Cotton:1899}\footnote{The Cotton is defined in $d>3$ via the (contracted) second Bianchi identity $\bnabla_a \bR^{a}{}_{bcd}=\bnabla_a \overline{W}^{a}{}_{bcd}+(d-3)\sC_{bcd}=0$.}
\beq
\sC_{abc}=2\bnabla_{[a}\bS_{b]c}
\eeq
where we used the convention $A_{[ab]}=\frac12 (A_{ab}-A_{ba})$ and we introduced the Schouten tensor, which in $3$ dimensions reads
\beq\label{sch}
\bS_{ab}=\bR_{ab}-\frac{\bR}{4} \bg_{ab}.
\eeq

The Cotton tensor is a vector-valued two-form. In $3$-dimensions, it can then be Hodge-dualized into what is sometimes called the Cotton-York tensor\footnote{The vanishing of this tensor is the equation of motion of the $3$-dimensional gravitational Chern-Simons action \cite{Deser:1982vy}.}
\beq\label{co}
\sC_{ab}=\ve_{acd}\sC^{cd}{}_b,
\eeq
where we introduced the Levi-Civita tensor $\ve_{acd}=\sqrt{|\bg|}\epsilon_{acd}$, with $\epsilon_{acd}$ the Levi-Civita symbol. In this paper, we work in Lorentzian signature, and thus the Hodge dual squares to $\star\star \eta=-(-1)^{p(3-p)} \eta$, for any $p$-form $\eta$. 

The Cotton-York tensor identically satisfies 
\beq
\sC_{ab}=\sC_{ba}\qquad Tr(\sC)=0\qquad \bnabla_a \sC^a{}_b=0,
\eeq
and therefore shares all the defining properties of the boundary stress tensor, on shell -- see \eqref{pi} and the discussion in \cite{Ciambelli:2018wre}. As mentioned, it is furthermore the $3$-dimensional conformal tensor: under the local rescaling $\bg_{ab}\to \frac{\bg_{ab}}{{\cal B}(x)^2}$, it transforms as
\beq
\sC_{ab}\to {\cal B}(x) \sC_{ab}.
\eeq

For the purpose of this paper, the most important property is that it vanishes if and only if  the metric is conformally flat
\beq\label{c0}
\sC_{ab}=0 \quad \Leftrightarrow \quad \bg_{ab}=e^{2\phi}\eta_{ab}.
\eeq
Consequently, this tensor instructs us about the conformal class of the boundary. In the literature, the vanishing (or not) of 
$\sC_{ab}$ 
at the boundary is sometimes referred as the bulk being asymptotically globally (or locally) AdS. We will see shortly that 
$\sC_{ab}$ 
plays a pivotal role in the definition of radiation in Anti de Sitter spaces, as expected since it is the boundary conformal tensor.

\paragraph{A Proposal for Radiation}

In the last part of this section, we review the  definition of radiation proposed by Fern\'andez-\'Alvarez and Senovilla \cite{Fernandez-Alvarez:2019kdd, Fernandez-Alvarez:2020hsv, Fernandez-Alvarez:2021yog, Fernandez-Alvarez:2021zmp, Fernandez-Alvarez:2021uvz, Senovilla:2022pym}, and study its possible enhancement to AdS spaces. The purpose is to have a geometric notion of radiation and a gauge-invariant condition allowing us to distinguish radiative from non-radiative solutions of the Einstein equations. We give here a brief and simplified treatment, using -- although not necessary -- the FG gauge as a constructive example. We refer to \cite{Senovilla:2022pym} for a more comprehensive analysis, and stress that we present these results from a physicist's viewpoint. This is also the approach we have with our proposal in AdS, which deserves a more thorough mathematical analysis in the future.

The first step is to conformally compactify the bulk by introducing the conformal factor $\Omega$: $\hat g_{\mu\nu}=\Omega^2 g_{\mu\nu}$, which in FG is simply $\Omega=\frac1r$. Then, we introduce the Bel-Robinson tensor \cite{Bel1, bel2}
\beq
\hat {\cal T}_{\mu\nu\rho\sigma}=\hat W^{\beta}{}_{\rho\mu\alpha} \hat W^\alpha{}_{\nu\sigma\beta}+*\hat W^\beta{}_{\rho\mu\alpha} *\hat W^\alpha_{\nu\sigma\beta},
\eeq
where $*W$ is the (bulk) Hodge dual of the (bulk) Weyl tensor. The finite order part of this tensor vanishes as we approach the boundary  \cite{Geroch:1977big}, but one obtains a non-zero limit by considering the rescaled tensor
\beq
\hat {\cal D}_{\mu\nu\rho\sigma}=\hat d^{\beta}{}_{\rho\mu\alpha} \hat d^\alpha{}_{\nu\sigma\beta}+*\hat d^\beta{}_{\rho\mu\alpha} *\hat d^\alpha_{\nu\sigma\beta} \qquad \text{with} \qquad \hat d^\mu{}_{\nu\rho\sigma}=\Omega^{-1} \hat W^\mu{}_{\nu\rho\sigma}.
\eeq

The normal to the compactified boundary is $\hat n=\hat n_\alpha \rd x^\alpha=\hat \nabla_\alpha \Omega \rd x^\alpha=-\frac{\rd r}{r^2}$, which is normalized to $1$ because $\hat g^{rr}=r^4$. Notice that this vector is by construction spacelike, as we are interested in radiation reaching the conformal boundary. Then, we define the canonical asymptotic supermomentum tensor
\beq\label{smt}
\hat{\cal P}^\mu=-\hat{\cal D}^\mu{}_{\alpha\beta\gamma}\hat n^\alpha \hat n^\beta \hat n^\gamma.
\eeq
In the absence of matter, it is a conserved vector on the boundary
\beq
\nabla_\mu\hat{\cal P}^\mu\stackrel{\cal B}{=}0.
\eeq
We are interested only in its boundary value and components tangent to the boundary, which in FG gauge amounts to considering $\hat{\cal P}^a$ at the boundary, called the super-Poynting vector. Another simplification that we do is to consider the presence or absence of radiation on the entire boundary, rather than a portion of it. Then, we propose the following characterization of radiation in AdS
\beq
\hat{\cal P}^a\stackrel{{\cal B}}{=}0 \quad \Leftrightarrow \quad \text{No Gravitational Radiation on} \ {\cal B}.
\eeq
For asymptotically de Sitter and flat spaces this is the result of \cite{Fernandez-Alvarez:2019kdd, Fernandez-Alvarez:2020hsv, Fernandez-Alvarez:2021yog, Fernandez-Alvarez:2021zmp, Fernandez-Alvarez:2021uvz, Senovilla:2022pym}. We have here extrapolated this definition to AdS spaces. We propose this criterion in AdS for the first time here, based on two main facts: First, we will see in the next section that in asymptotically flat spaces in Bondi gauge this condition reduces to the well-known condition that the news tensor must vanish on puncture-less spheres, and thus our AdS proposal matches this in the flat limit. Secondly, this criterion can be tested in known exact bulk solutions, and it gives rise to sensitive answers. For instance, we show in appendix \ref{appA} that this tensor does not vanish and captures radiation for Robinson-Trautmann spaces, which describe radiating black holes, whereas it vanishes for Taub NUT spaces, as expected since Taub NUT is not radiative. This is a positive test of the robustness of our criterion, since both mentioned spaces admit a non-vanishing Cotton, and thus we are here unveiling that the Cotton tensor alone is not enough to characterize radiation. We relegate to future investigations a more rigorous and mathematical study of this criterion in AdS, on the same lines as what done in dS and flat spaces in \cite{Fernandez-Alvarez:2019kdd, Fernandez-Alvarez:2020hsv}. In particular, a similar set of conditions has been derived in \cite{Holzegel:2015swa}, and we plan to make the connection more precise in the future.

Crucially for us, the super-Poynting vector can be rewritten in terms of the bulk electric and magnetic part of the Weyl tensor, which asymptote to the boundary stress tensor $\pi^{(0)}_{ab}$ and Cotton-York tensor $\sC_{ab}$, respectively.\footnote{This feature, and therefore the Cotton tensor, is of central relevance for AdS gravitational electric-magnetic duality, as discussed in \cite{deHaro:2007eg, deHaro:2007fg, Leigh:2007wf, Mansi:2008bs, Mansi:2008br}. See also \cite{Ciambelli:2020qny}.} Putting all these steps together, we arrive to
\beq\label{sp}
\hat {\cal P}^a\stackrel{{\cal B}}{=}2\sC_{b}{}^c \pi^{(0)}_{cd}\ve^{bda}=[\sC,\pi^{(0)}]_{bc}\ve^{bca}.
\eeq
This is the main result of this section: asymptotic radiation in AdS is encoded in the super-Poynting vector, which is the commutator between the boundary Cotton-York tensor and the boundary stress tensor. This implies that, in order to have radiation, they should both be non-vanishing. Since the Cotton-York tensor is a function of the boundary metric, we see that the shape of the latter is directly connected to the presence of radiation. This is in contrast to what happens in asymptotically flat spaces, where we will see that the radiation is encoded in subleading properties of the bulk metric as one approaches the boundary.

Note that, using the properties of the Levi-Civita symbol, and the definition \eqref{co}, we can rewrite the super-Poynting as
\beq
\hat\cP^a=-4 \sC^a{}_{bc}\pi^{(0)bc}.
\eeq
This not only makes explicit the parity properties, it also suggests using $\sC_{abc}$ to construct a geometric stress tensor, as we will do in Section \ref{sec:ccft}.

\subsection{AdS/CFT}\label{sec:adscft}

We are ready to explore the consequences of the non-radiative condition in the framework of the AdS/CFT correspondence \cite{Maldacena:1997re, Witten:1998qj}. We will adopt a geometric approach, and insist on properties of the bulk geometry rather than aspects of the dual field theory. The purpose is to study the repercussions of radiation and the presence/absence of the Cotton tensor at the boundary.

\paragraph{Conformal Isometries} The AdS/CFT correspondence is a duality between gravity in the bulk and a conformal field theory living on the boundary $\cal B$. The classic example of this duality stems from type IIB string theory on AdS$_5$ $\times$ S$^5$ but, for the purposes of this paper, we will discuss AdS$_4$ bulks, and content ourselves with Einstein gravity in the bulk. The duality is formulated from the action principle, stating that the bulk quantum gravity partition function is equal to the boundary CFT generating functional. The expectation value of the boundary stress tensor is then computed as the variation of the action with respect to the boundary metric, leading to the result that $\pi^{(0)}_{ab}$ is the boundary stress tensor, with its tracelessness and conservation 
\eqref{pi} corresponding to boundary conformal Ward identities. In fact, one needs to renormalize the stress tensor, which naively diverges due to the asymptotic nature of the boundary, see for instance \cite{Balasubramanian:1999re, Skenderis:2000in, deHaro:2000vlm, Skenderis:2002wp, Papadimitriou:2004ap, Papadimitriou:2005ii, Compere:2008us, Freidel:2008sh}, as well as more recent analysis on renormalization of the AdS symplectic structure at infinity \cite{Ruzziconi:2020wrb, Alessio:2020ioh, Fiorucci:2020xto, Ciambelli:2023ott, Grumiller:2016pqb, Campoleoni:2022wmf,  McNees:2023tus, McNees:2024}.

The boundary hosts a field theory on a fixed background. Therefore, the AdS/CFT dictionary is formulated imposing that the bulk metric approach a non-fluctuating structure at the conformal boundary, as originally studied in \cite{Henneaux:1985tv}. In other words, we require Dirichlet boundary conditions at the boundary $\delta \bg_{ab}=0$. Given \eqref{FG}, this implies
\beq
{\cal L}_{\xi} g_{ab}=O(r^0).
\eeq
Using the gauge preserving diffeomorphisms \eqref{adsxi}, it is a straightforward computation to show that the above condition reduces to
\beq
{\cal L}_Y \bg_{ab}=2\sigma \bg_{ab}.
\eeq
That is, the bulk diffeomorphisms preserving the FG gauge and Dirichlet boundary conditions are the boundary conformal isometries. This statement is true for an arbitrary boundary metric $\bg_{ab}$, but of course the solution to this equation depends on the specific structure of the boundary metric, which is an initial value problem for the radial reconstruction.

The maximal number of conformal isometries is reached for a boundary metric that is conformally flat $\bg_{ab}=e^{2\phi(x)}\eta_{ab}$, in which case we have that $Y^a$ generates the group $SO(2,3)$. These are also the isometries of Lorentzian AdS$_4$ spaces, or the asymptotic isometries of asymptotically globally AdS spaces. The starting point of the AdS/CFT duality is, from the geometric perspective, achieved imposing Dirichlet boundary conditions, and further requiring that the boundary metric is conformally flat. One can consider non-conformally-flat boundaries in AdS/CFT, but the asymptotic symmetry group is restricted. In this case the boundary hosts a CFT on a curved -- possibly time-dependent -- background. 

\paragraph{Cotton and Radiation} Now we can analyze how the geometric notion of radiation proposed in the previous section intertwines with AdS/CFT. The vanishing of the boundary Cotton-York tensor is a sufficient condition to guarantee the absence of radiation. Thanks to \eqref{c0}, it implies that the boundary metric is conformally flat, and thus that the asymptotic symmetry group with Dirichlet boundary conditions is $SO(2,3)$. Conversely, if the boundary Cotton-York tensor is not zero, then the boundary metric is not in the conformal orbit of the flat metric, and thus the boundary conformal isometries are strictly smaller than $SO(2,3)$. This is a theorem: a spacetime admits maximal number of conformal isometries if, and only if, it is conformally flat \cite{eisenhart1997riemannian}. Recalling \eqref{c0}, this property in $3$-dimensional spaces is encoded in the vanishing of the Cotton tensor such that one has
\beq
\sC_{ab} = 0\quad \Leftrightarrow \quad  \bg_{ab}=e^{2\phi}\eta_{ab} \quad \Leftrightarrow\quad \text{Max \# of Conformal Isometries} \quad \Rightarrow \quad \hat\cP^a\stackrel{{\cal B}}{=}0.
\eeq 
Putting these results together we see that the vanishing of the Cotton-York tensor must be required to have a CFT on flat space in the boundary in AdS/CFT, and in turns to have the maximal number of conformal isometries. But this implies the absence of background radiation. Under these conditions, one can still consider infinitesimal gravitons as stress tensor insertions in the boundary, but there are no radiative geometric fluxes arriving at the conformal boundary. If there were, the boundary metric would cease to be conformally flat, and there would be less global symmetries in the boundary field theory. As we show in the next section, this is a radical difference between AdS/CFT and flat space holography, where the asymptotic symmetry group is unaffected by the presence (or absence) of radiation, and so is the boundary Carrollian structure, i.e., the boundary degenerate metric and null generators. 

As mentioned, AdS backgrounds with non-vanishing boundary Cotton have been studied in the literature \cite{BernardideFreitas:2014eoi}, particularly in the context of the fluid/gravity correspondence 
\cite{Caldarelli:2011idw, Mukhopadhyay:2013gja}, and in relation to gravitational electric/magnetic duality \cite{Bakas:2008gz}. Exact Einstein solutions with radiation include the Robinson-Trautmann space, which describes an accelerated black hole \cite{Bakas:2014kfa, Skenderis:2017dnh, Ciambelli:2017wou, Adami:2024mtu}. The understanding of these spaces in AdS/CFT remains elusive, and it is currently under investigation. Furthermore, gravitational waves in holography have been studied in e.g. \cite{Polchinski:1999yd}. It is our intention to use our formalism to understand the "transient" phase in this setting.

The purpose of this section was to propose a notion of radiation in AdS that extends the known results for flat space. It is intimately related to the conformal tensor of the boundary spacetime, the Cotton-York tensor, which vanishes when considering dual CFT on flat spaces. Considering backgrounds with radiation in AdS/CFT  is an important challenge that must be addressed in the future, as it would allow us to contrast and compare better AdS/CFT and flat space holography.

In conclusion, in this section we have proposed a new way to characterize (non-perturbative) radiation in AdS, and we have shown that it requires the Cotton tensor of the boundary to be non-zero. This set-up is comprehended in the AdS/CFT correspondence, but it is more challenging to study. We argued that controlling this framework in AdS can have deep repercussions in the flat limit and the understanding of radiative degrees of freedom in holography in general.

\section{Flat}\label{sec:flat}

In this section we examine radiation in flat holography.   
In the first part, we discuss how, unlike in AdS, the presence of radiation does not restrict the asymptotic symmetries. After reviewing the $\Lambda$-BMS construction of~\cite{Compere:2019bua,Compere:2020lrt}, we show that our definition of radiation in AdS lands on the intrinsically flat definition in the $\Lambda\to 0$ limit. In the third part we connect these discussions to the celestial construction, showing how the $2$-dimensional Celestial stress tensor, supertranslation current, and $w_{1+\infty}$ generators can be extracted from the $3$-dimensional holographic stress tensor in the flat limit.

\subsection{Bondi Gauge and Radiation}\label{sec:bondi}

We now review the analogous setup regarding gauge fixing, identifying asymptotic symmetries, and defining radiation for asymptotically flat spacetimes.

\paragraph{Asymptotic Structure} A $4$-dimensional asymptotically flat spacetime (AFS) is a solution of the Einstein equations with vanishing cosmological constant
\beq\label{eq:EE}
G_{\mu\nu}=8\pi G_N T^M_{\mu\nu}.
\eeq

A convenient gauge for describing asymptotically flat spacetimes is Bondi gauge~\cite{Bondi:1962px,Sachs:1962wk,Sachs:1962zza}, where $r$ is a radial coordinate parameterizing outgoing (resp. incoming) geodesics near the null conformal boundary $\mathcal{I}^+$ ($\mathcal{I}^-$). We confine our attention to the analysis near future null infinity $\mathcal{I}^+$ in what follows. 
Using coordinates $x^\mu=(r,u,x^A)$ with $x^A=(z,\bz)$, the Bondi gauge metric is given by \cite{Bondi:1962px, Sachs:1962wk, Barnich:2011mi,Freidel:2021fxf}\footnote{See \cite{Campoleoni:2023fug} and \cite{Geiller:2022vto, Geiller:2024amx} for $4d$ relaxations of the Bondi gauge.}
\begin{equation}\label{fullBondiMetric}
    \rd s^2=e^{2\beta}\frac{V}{r}\rd u^2-2e^{2\beta}\rd u\rd r+g_{AB}(\rd x^A-U^A\rd u)(\rd x^B-U^B\rd u),
\end{equation}
where the Bondi gauge condition consists of fixing
\begin{equation}
    g_{rr}=0,\qquad g_{rA}=0,\qquad \p_r\det\left(\frac{g_{AB}}{r^2}\right) = 0.
\end{equation}
The conformal boundary is located at $r\rightarrow \infty$ holding $u=t-r$ fixed. In this limit the metric can be expanded as
\be\badat{3}\label{bondi}
\rd s^2=&-\frac{\bar R}{2}\rd u^2-2\rd u\rd r+r^2\gamma_{AB}\rd x^A \rd x^B+\frac{2m_B}{r}\rd u^2+rC_{AB}\rd x^A\rd x^B+D_B C^B{}_A \rd u\rd x^A\\
&+\frac{2}{3r}\left[N_A+C_{AB}D_CC^{CB}\right]\rd u\rd x^A+\dots
\eadat\ee
where the ${A,B,\dots}$ indices are raised and lowered using the metric $\gamma_{AB}$, $D_A$ is its Levi-Civita connection, and $\bar R$ its scalar curvature. The free data is given by the shear $C_{AB}(u,z,\bz)$
and subleading orders in $r$ are determined by solving the Einstein equations. In particular, after specifying the Bondi mass $m_B$ and angular momentum aspect $N_A$ at a particular cut $u=u_0$, their $u$-evolution is determined by the constraint equations $n^\mu [G_{\mu\nu}-8\pi G_NT^M_{\mu\nu}]|_{\mathcal{I}^+}=0$ where $n$ is the null normal of the conformal boundary:
\be\label{constraints}\badat{3}
\p_u m_B=&-\frac{1}{8}N_{AB}N^{AB}+\frac{1}{4}D_A D_B N^{AB}-4\pi G\lim\limits_{r\rightarrow\infty}r^2T^M_{uu},\\
\p_u N_A=& \ \pa_A m_B+\frac1{16}\pa_A (N_{BC}C^{BC})-\frac14 D_A C_{BC} N^{BC}-\frac14 D_B (C^{BC}N_{CA}-N^{BC}C_{CA})\\
&-\frac14 D_B(D^BD_C C^C{}_A-D_A D_C C^{BC})-8\pi G\lim\limits_{r\rightarrow\infty}r^2T^M_{uA}.
\eadat\ee
Here $N_{AB}=\p_u C_{AB}$ is the news tensor.

The infinitesimal symmetries $\xi(u,r,z,\bz)$ preserving the gauge conditions are given by
\be\badat{3}\label{xiyf}
\xi=& \ Y^A\pa_A+\frac{1}{2}D_AY^A(u\p_u-r\pa_r)+f\p_u+\frac12 D^AD_A f\p_r\\
&-\frac1r D^A (f+\frac12 u D_BY^B ) \pa_A+\frac{1}{4}D_AD^A D_BY^B u\pa_r+\dots
\eadat
\ee
where $f=f(z,\bz)$ and $Y=Y(z)$ correspond to supertranslations and superrotations, respectively. As we will show presently, these angle dependent enhancements of Poincar\'e to an infinite dimensional group are independent of the presence of radiation. This is in contrast to the AdS case considered above, whereby the symmetries~\eqref{adsxi} get restricted to the global conformal group (or further) once we fix the conformal class of the boundary metric $[\bg_{ab}(x)]$. Indeed the radiation in asymptotically flat spacetimes appears at subleading order in $r$ to the boundary metric, which is now degenerate. Namely, defining the conformally rescaled metric
\be\label{eq:cres}
\widetilde{\rd s}^2=\Omega^2 \rd s^2
\ee
with conformal factor $\Omega=\frac{1}{r}$, we have
\be\label{eq:carrflat}
\lim_{r\to\infty}\widetilde{\rd s}^2=\gamma_{AB} \ \rd x^A\rd x^B.
\ee
The metric on $\mathcal{I}^+$ (which is a $3$-dimensional manifold) is therefore degenerate, signaling the presence of a null manifold. 
A more detailed study of the intrinsic geometrical construction of such {\it Carrollian} geometries can be found in~\cite{Ciambelli:2018wre}, see \cite{Ciambelli:2023mir, Freidel:2024tpl} as well, and the description of null infinity in \cite{ Ashtekar:2024bpi}. The important feature is that this metric is not dynamical. Instead, what sources the evolution in equations \eqref{constraints} is the asymptotic shear $C_{AB}$, which appears at subleading order in the metric expansion. Indeed, the latter is going to dictate the radiative structure at infinity, whereas in AdS we saw that radiation is informed by the boundary Cotton tensor, and thus the boundary metric itself.

For our purposes, the flat limit of the AdS radiation vector will be made clearer by reformulating the AdS bulk geometry in a Bondi-like gauge. But first, let us introduce the analogous definition for radiation appropriate for $\Lambda=0$ spacetimes.

\paragraph{Radiation in Flat}

The main result of  \cite{Fernandez-Alvarez:2019kdd, Fernandez-Alvarez:2020hsv, Fernandez-Alvarez:2021yog, Fernandez-Alvarez:2021zmp, Fernandez-Alvarez:2021uvz, Senovilla:2022pym} is that the super momentum tensor \eqref{smt} vanishes asymptotically if and only if there is no radiation. In these references, it is shown that the leading order of the super momentum tensor evaluated on a cut of $\mathcal{I}^+$ has the form
\beq\label{pt}
\hat{\cal P}^r=2\pa_u N_{AB} \pa_u N^{AB}\qquad \hat{\cal P}^u= 2 D_C N^C{}_A D_B N^{AB}\qquad \hat{\cal P}^A=-4 \pa_u N^{AB}D_C N^C{}_B.
\eeq
Requiring $\hat{\cal P}^\mu\stackrel{{\cal B}}{=}0$  we then obtain that  $N_{AB}=N^{vac}_{AB}$, with $N^{vac}_{ab}$ defined in \cite{Compere:2018ylh}, such that, restricting to puncture-less spheres, we have
\beq\label{rad}
\hat{\cal P}^\mu\stackrel{{\cal B}}{=}0 \quad   \Leftrightarrow \quad N_{AB}=0.
\eeq
As we discuss below, one of the main results of this paper is to shown how the expressions \eqref{pt} can be recovered in the flat limit of the AdS notion of radiation. Note the important difference between the two cases: in AdS, absence of radiation on the boundary is achieved by demanding $\hat {\cal P}^a=0$, while in flat it is the whole super momentum tensor that must vanish, including $\hat {\cal P}^r=0$. We can understand why this comes about: geometrically, the timelike vector field at the boundary of AdS will be shown to contain both $\hat {\cal P}^r$ (in the leading piece) and $\hat {\cal P}^u$ (in the subleading piece) in the flat limit.

\subsection{Limit from AdS}\label{sec:ads2flat}

To facilitate comparing our AdS and flat definitions of radiation, we will now review the construction in~\cite{Compere:2019bua,Compere:2020lrt} of a Bondi-like gauge for asymptotically AdS spacetimes. There the authors consider solutions to~\eqref{eq:eeads} (focusing on vacuum solutions) in coordinates where $r$ is now an affine parameter for radial null geodesics and $u$ foliates the spacetime into null hypersurfaces. We can still use the Bondi gauge parameterization~\eqref{fullBondiMetric} but now solutions to~\eqref{eq:eeads} will have
\begin{align}
    \frac{V}{r} & =\frac{\Lambda}{3} r^2-r \ell-\frac{1}{2}\left(R[q]+\frac{\Lambda}{8} C_{A B} C^{A B}\right)+\frac{2 m_B}{r}+\mathcal{O}\left(\frac{1}{r^2}\right)\\
    \beta & =-\frac{1}{32 r^2} C^{A B} C_{A B}+\mathcal{O}\left(\frac{1}{r^3}\right) \\
    U^A & =-\frac{1}{2r^2} D_B C^{A B}-\frac{1}{r^3}\left(\frac{2}{3} N^A-\frac{1}{2} C^{A B} D^C C_{B C}\right)+\mathcal{O}\left(\frac{1}{r^4}\right) 
\end{align}
and
\begin{equation}\label{angularBondi}
      g_{A B}  =r^2 q_{A B}+r C_{A B}+D_{A B}+\frac{1}{r} \mathcal{E}_{A B} + \frac{1}{r^2} F_{A B}+\mathcal{O}\left(\frac{1}{r^3}\right) 
\end{equation}
where~$\ell=\p_u\ln\sqrt{q}$. Defining the conformally rescaled metric as in~\eqref{eq:cres}, the pullback at the AdS boundary $r\rightarrow\infty$
is
\be\label{eq:bndymet}
\widetilde{\rd s}^2|_{\mathcal{J}}=\frac{\Lambda}{3}\rd u^2+q_{AB}\rd x^A \rd x^B.
\ee
We indeed see that flat limit~\eqref{eq:carrflat} is a Carrollian limit of the boundary with $c^2=-\frac{\Lambda}{3} 
\rightarrow0$, where $c$ is the speed of light on the boundary. 

The main difference between the flat solution space and the AdS solution in Bondi gauge is that the asymptotic shear in AdS is a function of the boundary metric. Choosing for convenience the same boundary gauge fixing as in ~\cite{Compere:2019bua,Compere:2020lrt} we have
\begin{equation}\label{BondiCAB}
    \frac{\Lambda}{3}C_{AB}=(\p_u-\ell)q_{AB}.
\end{equation}
Therefore, if $\Lambda=0$, we find that $C_{AB}$ is free data. However, as soon as $\Lambda\neq0$ the shear is determined by the leading metric $q_{AB}$ which, moreover, must have a non-trivial time dependence if we want to have radiation in the flat limit. 

To compare what this means in terms of our discussion in the previous section one should go back to Fefferman-Graham gauge, which was conveniently also done in~\cite{Compere:2019bua,Compere:2020lrt}. One obtains
\begin{equation}\label{FGdata}
    \bg_{ab} = \begin{pmatrix}
        \frac{\Lambda}{3}&0\\
        0&q_{AB}
     \end{pmatrix}, \qquad
     \pi^{(0)}_{ab}=
    \begin{pmatrix}
        -\frac43 M^{(\Lambda)} & -\frac23 N_B^{(\Lambda)}\\
        -\frac23 N_A^{(\Lambda)} & J_{AB}+\frac{2}{\Lambda}M^{(\Lambda)}q_{AB}
    \end{pmatrix}
\end{equation}
where
\begin{align}\label{mnj0}
    M^{(\Lambda)}&=m_B+\frac{1}{16}\left(\partial_u+\ell\right) (C_{CD}C^{CD})\\
    N_A^{(\Lambda)}&=N_A-\frac{3}{2\Lambda}D^B\left(N_{AB}-\frac12\ell C_{AB}\right) - \frac34\partial_A\left(\frac{1}{\Lambda}R[q]-\frac38C_{CD}C^{CD}\right)\\
    J_{AB}&=-\mathcal{E}_{AB} - \frac{3}{\Lambda^2}\partial_u\left( N_{AB}-\frac12\ell C_{AB}\right) + \frac{3}{2\Lambda}q_{AB}C^{CD} \left(N_{CD}-\frac12\ell C_{CD}\right)\label{mnj1} \\
    &\qquad +\frac{3}{\Lambda^2}\left(D_AD_B-\frac12q_{AB}D_CD^C\right)\ell - \frac{1}{\Lambda}\left(D_{(A}D^CC_{B)C}-\frac12q_{AB}D^CD^DC_{CD}\right) \nonumber\\
    &\qquad + C_{AB}\left[\frac{5}{16}C_{CD}C^{CD} + \frac{1}{2\Lambda}R[q]\right]. \nonumber
\end{align}
Comparing with our discussion in section~\ref{sec:adscft}, our boundary metric displayed in equation ~\eqref{FGdata} implies that our radiative spacetimes with $\p_u q_{AB}\neq 0$ will generically have a smaller conformal isometry group. 

Meanwhile the second equation in~\eqref{FGdata} demonstrates the breakdown of the Fefferman-Graham gauge in the flat limit, since the angular components of the stress tensor are singular as $\Lambda\rightarrow 0$. Nevertheless, the singular terms in the flat limit of the conservation equation 
$\nabla_aT^{a}{}_{b}=0$ for
\be\label{3d stress}
T_{ab}=\frac{\sqrt{3|\Lambda|}}{16\pi G_N}\pi^{(0)}_{ab}
\ee
consistently impose the constraint equations for asymptotically flat spacetimes~\eqref{constraints}.

\paragraph{Matching Radiation}

We now wish to take the flat limit of \eqref{sp}. The Cotton tensor has a complicated expression in terms of $3$-derivatives of the boundary metric. We split the boundary metric into a leading, time-independent, piece plus subleading orders in $\Lambda$
\beq\label{eq:qexp}
q_{AB} = \gamma_{AB}+\sum_{n=1}^{\infty}\Lambda^n q_{AB}^{(n)},
\eeq
such that \eqref{BondiCAB} can be conveniently solved. 

For simplicity, we assume that $\gamma_{AB}$ is the metric of the $2$-sphere, with $\bar R=2$. Then, we get 
\beq\label{Cab}
C_{AB}=3\left(\p_u q^{(1)}_{AB}-\frac12\gamma^{CD}\pa_u q^{(1)}_{CD}\gamma_{AB}\right)+{\cal O}(\Lambda).
\eeq
We can furthermore compute the leading orders of (\ref{mnj0}-\ref{mnj1})\footnote{The covariant derivative $D_A$ is here the Levi-Civita connection of $\gamma$, which is only the leading order of the covariant derivative $D_A$ of the previous subsection, in the flat limit. However, since we are interested in the first non-vanishing order, we will not make distinction between the two, which is clear from context.}
\beq
& M^{(\Lambda)}=m_B+\frac1{16}\pa_u(C_{AB}C^{AB})+{\cal O}(\Lambda) &\\
& N_A^{(\Lambda)}=-\frac3{2\Lambda}D^B N_{AB}+{\cal O}(\Lambda^0) & \\ 
& J_{AB}=-\frac3{\Lambda^2}\pa_u N_{AB}+{\cal O}(\Lambda^{-1}) &
\eeq
To leading order, the spatial indices are raised and lowered with $\gamma_{AB}$. We can now compute the Schouten tensor \eqref{sch} to leading order
\beq\label{eq:schouten}
\bS_{ab}=\begin{pmatrix}
    -\frac{\Lambda}{4}\gamma^{MN}\pa_u^2 q^{(1)}_{MN}-\frac{\Lambda}{6}+{\cal O}(\Lambda^2) & \frac{\Lambda}{2}(D^D\pa_u q^{(1)}_{BD}-D_B(\gamma^{MN}\pa_u q^{(1)}_{MN}))+{\cal O}(\Lambda^2)\\
    \frac{\Lambda}{2}(D^D\pa_u q^{(1)}_{AD}-D_A(\gamma^{MN}\pa_u q^{(1)}_{MN}))+{\cal O}(\Lambda^2) & -\frac12N_{AB}+{\cal O}(\Lambda)
\end{pmatrix}
\eeq
Notice that, remarkably, the Schouten tensor admits a smooth flat limit, which is simply the news tensor. There is a further condition imposed in~\cite{Compere:2019bua}, where the authors gauge-fix the area of the transverse space. This is allowed, exploiting the Weyl rescaling of the boundary. We can use this to impose
\be\label{detcond}
\gamma^{AB}\pa_u q^{(1)}_{AB}=0,
\ee
such that $C_{AB}=3\pa_u q^{(1)}_{AB}+{\cal O}(\Lambda)$. We will require this in section \ref{sec:adslift}, while for the remainder of this section, for the sake of generality, we will compute the radiation vector without this restriction.

The Levi-Civita tensor is given by
\beq
\ve_{abc}=\sqrt{|\bg |}\epsilon_{abc}=\sqrt{\frac{|\Lambda|}{3}}\sqrt{q}\epsilon_{abc}=\sqrt{\frac{|\Lambda|}{3}}\sqrt{\gamma}\epsilon_{abc}+{\cal O}(\Lambda^{3/2}).
\eeq
We furthermore introduce the $2$-dimensional Levi-Civita symbol define via $\epsilon_{uAB}=\epsilon_{AB}$, such that
\beq
\ve_{uAB}=\sqrt{\frac{|\Lambda|}{3}}\sqrt{\gamma}\epsilon_{uAB}+{\cal O}(\Lambda^{3/2})=\sqrt{\frac{|\Lambda|}{3}}\sqrt{\gamma}\epsilon_{AB}+{\cal O}(\Lambda^{3/2})=\sqrt{\frac{|\Lambda|}{3}}\ve_{AB}+{\cal O}(\Lambda^{3/2}),
\eeq
where in the last equality we've defined $\ve_{AB}=\sqrt{\gamma}\epsilon_{AB}$. We remark that, given that we are in Lorentzian signature, we have
\beq
\ve^{abc}=g^{ad}g^{be}g^{cf}\ve_{def}=\frac{-1}{\sqrt{|\bg|}}\epsilon^{abc}=-\sqrt{\frac{3}{|\Lambda|}}\frac{1}{\sqrt{q}}\epsilon^{abc},
\eeq
such that $\ve^{abc}\ve_{abc}=-6$. Using this and \eqref{Cab}, we get
\beq\label{cotton}
\sC_{ab}=\begin{pmatrix}
    -\frac{|\Lambda|^{3/2}}{\sqrt{3}}\ve^{AB}\left(D_A D^D\pa_u q^{(1)}_{BD}+\frac12 N_B{}^C\pa_u q^{(1)}_{AC}\right)+{\cal O}(\Lambda^{5/2}) & -\sqrt{\frac{|\Lambda|}{3}}\ve^{CD}D_CN_{DB}+{\cal O}(\Lambda^{3/2})\\
    -\sqrt{\frac{|\Lambda|}{3}}\ve^{CD}D_CN_{DA}+{\cal O}(\Lambda^{3/2}) & -\sqrt{\frac{3}{|\Lambda|}}\ve_{AC}\pa_uN^{C}{}_{B}+{\cal O}(\Lambda^{1/2})
\end{pmatrix}~.
\eeq
We can use these expansions to evaluate the flat limit of the super-Poynting vector, defined in \eqref{sp}. One finds
\beq
\hat\cP^u&\stackrel{\cal B}{=}&2\left(\frac{3}{\Lambda}\right)^3\pa_u N^{AB}\pa_u N_{AB}+\left(\frac{3}{\Lambda}\right)^2 \hat\cP^u_{(-2)}+{\cal O}(\Lambda^{-1})\\
\hat\cP^A&\stackrel{\cal B}{=}&2\left(\frac{3}{\Lambda}\right)^2\pa_u N^{AB}D_C N^C{}_{B}+{\cal O}(\Lambda^{-1}),
\eeq
where we've used $\hat\cP^u_{(-2)}$ to indicate the subleading order in the expansion. Note that, to leading order in $\Lambda$, $\hat \cP^a$ is determined entirely by $\hat \cP^u$, and thus it is timelike.

This is the main result of this section, and shows how the leading order of the super-Poynting gives rise to radiation in the asymptotically flat spaces. The full set of radiation conditions~\eqref{pt} can be attained by setting order by order this vector to zero. Requiring the leading order of $\hat\cP^u$ to vanish gives $\pa_u N^{AB}\pa_u N_{AB}=0$ and thus $\pa_u N_{AB}=0$. This trivializes the leading order of $\hat\cP^A$, and a further computation shows that under this condition we have\footnote{For any traceless tensor $H_{AB}$, one has $-2H_{CD}D^{[D}N^{A]C}=N_B{}^A D_C N^{CB}$.}
\beq\label{Pusub}
\hat\cP^u_{(-2)}\stackrel{\cal B}{=}\frac43 D_B N^{BC}D_A N^A{}_C.
\eeq
This means that setting the first two orders in $\hat\cP^u$ and the leading order in $\hat\cP^A$ to zero in the flat limit is exactly equal to solving for \eqref{rad}, given \eqref{pt}.

In conclusion, we have seen that the flat limit of the AdS radiative condition gives exactly the flat radiative condition. The main difference, as we stressed above, is that radiation in AdS impacts the asymptotic symmetries, while the latter are independent of the presence/absence of radiation for asymptotically flat spaces. En passant, we have also found how the radiative data are encoded in the Schouten and Cotton tensor. 
 
\subsection{Celestial Holography}\label{sec:ccft}

Celestial Holography proposes that gravitational scattering in asymptotically flat spacetimes is dual to a CFT living on the celestial sphere \cite{Strominger:2017zoo,Pasterski:2019ceq,Raclariu:2021zjz,Pasterski:2021rjz,Pasterski:2023ikd}. While this proposal was originally motivated by following a bottom up approach, starting from the asymptotic symmetry group, there have been recent top down successes using tools from twisted holography~\cite{Costello:2022wso,Costello:2022jpg}. Indeed the superrotations discussed above imply a Virasoro symmetry~\cite{Kapec:2014opa} that promises to bring $2d$ CFT techniques to bear on $4d$ physics. However, the codimension-2 nature of this dual makes this proposal a bit exotic compared to its AdS counterpart. 

After reviewing the origins and definitions of the celestial symmetries in section~\ref{sec:celestialsym}, we will show how to extract the respective currents from a dimensional reduction of the flat limit of the $3d$ boundary geometry in radiative AdS spacetimes in section~\ref{sec:adslift}.
Two relevant references that also give prescriptions for extracting these modes from flat limits of AdS are~\cite{deGioia:2023cbd} and~\cite{Hijano:2019qmi,Hijano:2020szl}, respectively. We will comment on these below, however it is worth mentioning from the start the important distinction that, as in the previous sections, we will be considering {\it radiative} AdS spacetimes.

\subsubsection{Celestial Symmetry Generators}\label{sec:celestialsym}

The Celestial CFT proposal was motivated by the fact that the BMS group implies an enhancement of Lorentz invariance to a Virasoro symmetry~\cite{Kapec:2014opa,Kapec:2016jld,Strominger:2017zoo}. Taking this 2d description seriously led Strominger and collaborators~\cite{Strominger:2021mtt,Guevara:2021abz} to identify a further tower of $w_{1+\infty}$ generators. Let us review how the BMS Ward identities follow from soft limits, and $w_{1+\infty}$ generators arise from collinear limits.

\paragraph{BMS Ward Identities in Celestial CFT} As discussed above, asymptotically flat spacetimes obey a BMS symmetry generated by~\eqref{xiyf}. The canonical charges can be expressed in terms of the Bondi mass and angular momentum aspect. Treating $\mathcal{I}^+$ as a Cauchy surface for the out state, we have~\cite{Barnich:2011mi, Pasterski:2019ceq} 
\be\label{Qfy}
Q^+[\xi]=\frac{1}{8\pi G}\int_{\mathcal{I}^+_-} [(2f+uD_AY^A)m_B + Y^A N_A].
\ee 
Now the $u$-evolution of the Bondi mass and angular momentum aspects are constrained by~\eqref{constraints}. We can use this to relate charges at the past cut of future null infinity $\mathcal{I}^+_-$ to fluxes along null infinity and data from any remaining massive fields at $i^+$ that contribute to the Bondi mass and angular momentum aspect at $\mathcal{I}^+_+$. For example, for supertranslations
\be
4\pi GQ^+[\xi_f]=\int_{\mathcal{I}^+} \hspace{-.1em} [-\frac{1}{4}D_A D_B N^{AB}+T_{uu}]f-\int_{\mathcal{I}^+_+} \hspace{-.1em} m_B f.
\ee
Here we have grouped terms quadratic in the shear into a shear inclusive ANEC. We can split this charge into a soft and hard part
\be\label{soft_hard}
Q^+[\xi]=Q_S^+[\xi]+Q_H^+[\xi],
\ee
where the soft part is linear in the radiative field and the hard part includes the massive matter and ANEC terms. Though the constraint equations~\eqref{constraints} imply these charges are not conserved as a function of the time $u$ of our different cuts, there is still a symmetry of the $\cal S$-matrix! 

A key insight of Strominger~\cite{Strominger:2013jfa,Strominger:2013lka} was to identify a diagonal subgroup of $BMS^+\times BMS^-$ whose charges at $\mathcal{I}^+_-$ and $\mathcal{I}^-_+$ can be identified via an antipodal matching condition of the fields near spatial infinity. Inserting this relation into perturbative $\cal S$-matrix elements 
\be\label{wardid}
\langle out | Q^+[\xi] \mathcal{S}-\mathcal{S}Q^-[\xi]|in\rangle=0,
\ee
gives a Ward identity~\cite{He:2014laa,Kapec:2014opa} as a consequence of the leading~\cite{Weinberg:1965nx} and subleading~\cite{Cachazo:2014fwa} soft graviton theorems. The Celestial Holography program has its origins in further identifying these $4d$ asymptotic symmetry identities as $2d$ conformal Ward identities. For particular choices of the supertranslation and superrotation parameters the soft operators become holomorphic currents~\cite{Strominger:2013jfa,Kapec:2016jld}
\be\label{bms_currents}
P_z=\frac{1}{4G}D^z \int du N_{zz}, ~~~
T_{zz} = \frac{i}{8\pi G_N} \int d^2w \frac{1}{z - w} D_w^2 \bar{D}^{\bw} \int du u N_{\bar{w}\bar{w}}
\ee
which physically measure the leading and spin memory effects~\cite{Strominger:2014pwa,Pasterski:2015tva}.
When inserted into $\cal S$-matrix elements the first takes the form of a Kac-Moody current, while the latter gives a $2d$ stress tensor Ward identity
\be\label{Tward}
\langle T_{zz}\mathcal{O}_1...\mathcal{O}_n\rangle=\sum_k\left[\frac{h_{k}}{(z-z_k)^2}+\frac{D_{z_k}}{z-z_k}\right]\langle\mathcal{O}_1...\mathcal{O}_n\rangle
\ee
 where
\be 
h_k=\frac{1}{2}\left(\ell_k-\omega_k\p_{\omega_k}\right),~~~\bar{h}_k=\frac{1}{2}\left(-\ell_k-\omega_k\p_{\omega_k}\right),
\ee
motivating the switch to a boost eigenbasis to diagonalize the weights. 

\paragraph{w$_{1+\infty}$ Symmetries of Self Dual Gravity} While the soft currents corresponding to the BMS generators were identified after manipulating the known covariant phase space representations of the BMS group, the $2d$ CCFT framework led to surprising additional symmetries. To make the codimension-2 hologram consistent with the number of on-shell degrees of freedom, we need a continuous spectrum of celestial weights for each bulk field~\cite{deBoer:2003vf,Pasterski:2017kqt}. Indeed, representing the amplitude in energy, position, or weight variables is just a change of basis~\cite{Donnay:2022sdg}
\begin{align}\label{soft currents}
    \mathcal{O}_{\Delta,2}^+(z,\Bar{z}) = \frac{\Gamma(\Delta-2)}{4\pi i^{\Delta+1}} \int_{-\infty}^\infty du u^{-\Delta+2}\p_uN_{zz}.
\end{align} 
We can analytically continue these weights to negative integer values where the $SL(2,\mathbb{C})$ multiplets shorten. Namely, defining the residues at these weights gives a set of holomorphic currents
\begin{equation}\label{eq:resH}
    H^k(z,\bz) ~:=~ \lim_{\epsilon\to 0}\,\epsilon\,{\cal O}_{k+\epsilon,2}(z,\bz)=\sum_{m=\frac{k-2}{2}}^{\frac{2-k}{2}}\,\bz^{-\frac{k-2}{2}-m}\,H^k_m(z)
\end{equation}
Upon examining the collinear limits of scattering amplitudes we see that the operator product expansion of these modes~\cite{Fan:2019emx,Fotopoulos:2019vac,Pate:2019lpp} closes and the radial commutator 
\be\label{rcom}
[A,B](z)=\frac{1}{2\pi i}\oint_z dw A(w)B(z)
\ee
furnishes a representation of  $ \wedge L  w_{1+\infty}$ symmetry algebra~\cite{Guevara:2021abz,Strominger:2021mtt}
\begin{equation}
    \Big[w^p_n, w^q_m\Big] ~=~ \Big[ n(q-1) - m(p-1)\Big]\,w^{p+q-2}_{m+n},~~~~   w^p_n ~=~ \frac{1}{\kappa}\,(p-n-1)!(p+n-1)!\,H_n^{-2p+4}~.
    \label{eq:wloop}
\end{equation}
Here $p=1,\frac{3}{2},2,\frac{5}{2}...$ and in this derivation we restrict to the wedge $1-p\le m\le p-1$, though there are interesting extensions beyond the wedge~\cite{Freidel:2021ytz,Freidel:2023gue,Hu:2022txx,Hu:2023geb}. While these symmetries were extracted in a bit of an exotic way from the point of view of the original amplitudes, they find a natural interpretation in twistor space as symmetries of self-dual gravity~\cite{Adamo:2021lrv}. Surprisingly, the convergence of tools from soft physics, twistor theory, and twisted holography around this common symmetry group led to the first top down construction of a celestial duality~\cite{Costello:2022jpg}!

\subsubsection{Lifting Celestial Currents to Radiative AdS}\label{sec:adslift}

The goal of the remainder of this section is to identify how to uplift these celestial currents to radiative AdS spacetimes.

\paragraph{Extracting Radiation from the Flat Limit} 
We have seen the news tensor appear in various places in our discussions of the boundary geometry for radiative AdS spacetimes in $\Lambda$-BMS gauge. In~\eqref{eq:schouten} we saw the Schouten tensor~\eqref{sch} limited to the news  
\beq\label{eq:schoutenlimit}
\lim_{\Lambda\rightarrow 0}\bS_{ab}=\begin{pmatrix}
    0 & 0\\
   0 & -\frac12N_{AB}
\end{pmatrix}.
\eeq
Meanwhile the leading singular components of the boundary stress tensor~\eqref{FGdata}-\eqref{3d stress} and Cotton tensor~\eqref{cotton} are proportional to it's $u$-derivative
\beq
\lim_{\Lambda\rightarrow0}\left(\frac{|\Lambda|}{3}\right)^{\frac{3}{2}}16\pi G_N T_{ab}=\begin{pmatrix}
  0 & 0\\
    0 & -\pa_uN_{AB}
\end{pmatrix},~~~\lim_{\Lambda\rightarrow0}\sqrt{\frac{|\Lambda|}{3}}\sC_{ab}=\begin{pmatrix}
  0 & 0\\
    0 & -\ve_{AC}\pa_uN^{C}{}_{B}
\end{pmatrix}~.
\eeq
Indeed the fact these are both proportional to $\p_u N_{AB}$ is what gave rise to the simplification in extracting the form of the subleading term in the super-Poynting vector~\eqref{Pusub} for non-radiative solutions.

From~\eqref{soft currents} we thus see how to extract the celestial primary gravitons from any of these $3d$ geometric objects.
The celestial currents arise from the conformally soft modes that appear as residues at integer dimensions. Namely, the BMS supertranslation current and celestial stress tensor can be extracted from the first two $u$-moments via~\eqref{bms_currents}.  Meanwhile, the $w_{1+\infty}$ generators can be extracted using~\eqref{eq:resH} and projecting onto the holomorphic currents by the appropriate contour integrals on the complexified celestial sphere.

Let us briefly compare this prescription to other precedents for extracting the soft currents from flat limits of AdS/CFT. Our method is reminiscent of the AdS extrapolate set-up discussed for scalars in~\cite{Hijano:2019qmi} and for photons in~\cite{Hijano:2020szl}. The gravitational case was considered in~\cite{deGioia:2023cbd}, where the conservation of the stress tensor and how that constrains its OPE, and that of its shadow transform, was all that was needed  to get a proposed soft theorem. In particular, they identify the first two orders in the $u\rightarrow0$ ($\tau=\frac{\pi}{2}+\frac{u}{R}$) of the angular components of the ($3d$) shadow stress tensor as the leading and subleading soft graviton.  This is seemingly different from our proposal, where we can extract the news directly from the most singular part of the stress tensor without any shadow transform. However, we again have to be careful about the modified falloffs for our radiative spacetimes when comparing proposals. Namely, we are considering a different class of boundary conditions for our radiative AdS spacetimes.

\paragraph{Recovering the Constraint Equations in the Flat Limit}
In our review of the BMS Ward Identities, we saw that the form of the soft theorem in flat spacetimes follows from the constraint equations. Namely, taking $u$ integrals of the Bondi mass and angular momentum aspect evolution equations and antipodally matching the in and out charges.\footnote{A better understanding of the origin of the antipodal matching condition from the perspective of operators at $\tau=\pm\frac{\pi}{2}$ scattering to the bulk point for radiative AdS spacetimes is worth pursuing. See~\cite{Hijano:2020szl} for an explicit treatment of both the in and out modes for the photon Ward identity perturbing around global AdS.}
From the point of view of the soft operators at future null infinity we have focused on herein, the constraint equations imply how the primary descendants of the conformally soft currents are sourced~\cite{Pasterski:2021fjn,Pasterski:2021dqe} (see~\cite{Freidel:2021dfs} for the Weyl-BMS analog). The form of the soft theorem is then just the Green's function for the differential operators taking us to the primary descendant.

As we alluded to in our discussion of $\Lambda$-BMS gauge above, and demonstrated in~\cite{Compere:2019bua}, these constraint equations can be recovered from the flat limit of the conservation law $\bnabla^a T_{ab}=0$.\footnote{This was also established in the flat limit of the fluid/gravity correspondence in \cite{Ciambelli:2018wre}, see also \cite{Campoleoni:2022wmf} for $3d$ bulks and \cite{Campoleoni:2023fug}.}  More specifically, assuming from now that \eqref{detcond} holds, 
\renewcommand{\arraystretch}{1.5}
\begin{equation}\label{summarize constraints}
    \begin{array}{ccc}
        \bnabla^aT_{a u}=0 & \Longleftrightarrow &  G_{uu}=0, \\
        \bnabla^aT_{a z}=0  & \Longleftrightarrow & G_{uz}=0.
    \end{array}
\end{equation}
However, while the conservation equations have a well defined flat limit, the stress tensor itself does not. Let's elaborate on this point.  

What we want to do is expand~\eqref{summarize constraints} in powers of $\Lambda$. If we expand the boundary metric as in~\eqref{eq:qexp},
we find that the connection for this metric is regular in the $\Lambda\rightarrow0$ limit while the inverse metric carries powers of $\frac{1}{\Lambda}$. As such it will be convenient to look at the raised index object $T^a{}_{b}$. 
When evaluating $\bnabla_a T^{a}{}_b=0$  the singular terms come from the first two orders
\be\label{tsing}
(T^{\rm sing})^a{}_{b}=\frac{1}{|\Lambda|^{\frac{3}{2}}}{T^{(-3/2)}}^a{}_{b}+\frac{1}{|\Lambda|^\frac{1}{2}}{T^{(-1/2)}}^a{}_{b} + ~{\rm reg.}
\ee
It is straightforward to extract the leading singular term from~\eqref{FGdata}-\eqref{3d stress}
\be\label{tn2}
{T^{(-3/2)}}^a{}_{b}= \frac{9}{16\pi G_N} \begin{pmatrix}
        0 &  D_C N^C{}_{B}\\
       0 & -\partial_uN^{A}{}_B
    \end{pmatrix} .
\ee
The fact that the conservation law holds for all orders in the $\Lambda$ expansion implies that the leading singular term will itself be a conserved tensor. However the leading and subleading terms in $T^a{}_b$ will mix due to the subleading terms in the connection. For example, the Bondi mass loss formula follows from a subleading term in the connection combining with the $AB$ components of~\eqref{tn2}.

On the one hand, we have an interesting picture that it's the part of the stress tensor that is singular in the $\Lambda\rightarrow0$ limit that contributes to the constraint equations of the asymptotically flat spacetimes. This is seemingly quite different from the discussion of flat limits in~\cite{deGioia:2023cbd}. On the other, the mixing of orders prevents us from extracting a conserved stress tensor and connection defined for the Carrollian theory at $\Lambda=0$. If we let ourselves reshuffle the conservation equations into sources on the right hand side (as done from an intrinsically Carrollian perspective in~\cite{Donnay:2022aba,Donnay:2022wvx}) we have another trick at our disposal. So far we have only expanded in small $\Lambda$, keeping all orders in $G_N$. If we additionally expand in $G_N$ we have the nice feature that the terms that were mixing orders in $\Lambda$ are quadratic. In particular, the linearized part of ${T^{(-1/2)}}^a{}_{b}$ will be conserved with respect to the non radiative connection $(\p_u,D_A)$ up to terms that are quadratic in the shear. Grouping these as sources and matter stress tensor contributions 
recasts the $3d$ stress tensor conservation law into a form reminiscent of the soft and hard splitting of~\eqref{soft_hard}.

\paragraph{A Regularized Stress Tensor for Self-Dual Solutions}  

A perhaps more geometric way to see the obstruction to defining a conserved $3d$ stress tensor in the flat limit is to study the interplay between the leading divergent mode \eqref{tn2} and those of the  Cotton tensor. Introduce the one-form normal to $u$-constant cuts
\beq
n=\rd u,
\eeq
one can check that the combination
\be
\hat{T}_{ab}=T_{ab} -\frac{\alpha}{2} n^c\ve_{cbd}\sC^d{}_{a}=T_{ab} +\alpha n^c\sC_{cab}\qquad \text{with}\qquad \alpha=\sqrt{\frac{3}{|\Lambda|}}\frac{1}{8\pi G_N}=\frac{\ell_{AdS}}{8\pi G_N}
\ee
has all components of $\hat{T}^a{}_{b}$ of $\mathcal{O}(|\Lambda|^{-1/2})$ in the flat limit. Therefore, in this combination the divergences are exactly cancelled. Note that, contrary to the combination of $T$ and $C$ coming from bulk gravitational electro-magnetic duality \cite{Leigh:2007wf,  Mansi:2008bs, Mansi:2008br, Caldarelli:2011idw, Mukhopadhyay:2013gja, Bakas:2008gz, 
 Ciambelli:2020qny}, our resulting tensor is parity even but is not symmetric, which is consistent with the symmetries of a Carrollian stress tensor \cite{Ciambelli:2018ojf}. While the Cotton tensor is identically conserved, one generically has that $\bnabla_a n^c$ does not vanish, and instead is proportional to the shear
\beqn
\bnabla_a\hat{T}^{a}{}_{b} = -\frac{\alpha}{2}\ve_{cbd}\sC^{da}\bnabla_a n^c=-\frac{3\alpha}{2\Lambda}\ve_{cbd}\sC^{da}\Gamma^c_{au},
\eeqn
such that
\beqn
\bnabla_a\hat{T}^{a}{}_{u}=\frac{\alpha}{4}\sqrt{\frac{3}{|\Lambda|}}\pa_u N^{AC}C_{AC}+{\cal O}(\Lambda^0),\qquad
\bnabla_a\hat{T}^{a}{}_{B}=\frac{\alpha}{4}\sqrt{\frac{3}{|\Lambda|}}D_{C}N^{CA} C_{AB}+{\cal O}(\Lambda^0).
\eeqn
Remarkably, we observe that the right hand sides vanish for (anti-)self dual spacetimes!
Namely, we have a $3d$ tensor built from a combination of the boundary stress tensor and Cotton tensor that: 1. has a well defined flat limit, 2. is conserved for self dual solutions, and thus 3. reproduces the appropriate constraint equations. Given the central role of self dual gravity to the celestial $w_{1+\infty}$ symmetry, this seems like an interesting conserved current in this set-up, and deserves further study. 

\section{Discussion}\label{sec:conclusions}

In this manuscript, we have provided a geometric definition of radiation in AdS spacetimes. Working in $4d$ bulks, we have shown that it is intimately tied with the $3d$ boundary Cotton tensor. The AdS/CFT framework requires us to work with Henneaux-Teitelboim boundary conditions \cite{Henneaux:1985tv}, and exploit that, when the boundary metric is conformally flat, the asymptotic symmetry group is the conformal group at the boundary. The boundary Cotton must be non-vanishing to have radiation in AdS, which in turns implies that the boundary is not conformally flat and, in turn, that the conformal group is smaller than $SO(2,3)$. This posits an interesting challenge in holography: how do we include in holography spacetimes with radiation? In other words, how do we account for a non-vanishing Cotton tensor in the boundary CFT? While we have only formulated these far-reaching questions, what we have established is that these must be addressed to connect AdS/CFT to celestial holography, where it is well-known -- though remarkable when compared to its AdS counterpart -- that the asymptotic symmetry group is not restricted due to the presence of radiation. 

The main results of our work are two-fold. First, we have matched in the flat limit of the $\Lambda$-BMS construction the AdS geometric notion of radiation with the usual one, that is, that the news tensor must vanish on puncture-less spheres. Secondly, we have identified the tensors admitting a smooth flat limit and giving rise to the radiative tensors. These are the Schouten tensor, and an interesting combination of the stress tensor and Cotton tensor, that deserves further study. To reiterate, we have shown how to provide a geometric definition of radiation in AdS and used this to relate the celestial currents to their AdS counterparts. Let us close with some outlooks.

\paragraph{Perturbations around Global AdS}
We saw in section~\ref{sec:ads} that the presence of radiation reduced the asymptotic symmetry group of the AdS hologram. Namely,  background radiation in AdS $\rightarrow$ non-zero Cotton tensor $\rightarrow$ smaller asymptotic symmetry group. The stark contrast with the asymptotically flat case 
makes it worth clarifying the sense in which we also expect to be able to reproduce gravitational $\cal S$-matrix elements from AdS/CFT~\cite{Penedones:2010ue,Gary:2009ae,Fitzpatrick:2010zm,Komatsu:2020sag,Hijano:2019qmi, Banerjee:2022oll}. Our discussion of the phase space for either asymptotically AdS or asymptotically flat bulks made no restriction on the magnitude of the radiation. If we restrict ourselves to perturbative gravity around global AdS we are back to the usual conformal isometry group. By contrast, the asymptotically flat spacetimes have an enhanced symmetry algebra, perturbative in $G_N$ or not. While isolating the linear and nonlinear parts let us perform a hard and soft splitting of the Carrollian stress tensor in section~\ref{sec:flat}, it wasn't essential to that construction. Instead, in the $\Lambda$-BMS construction it was the manner in which we were perturbative in $\Lambda$ -- rather than $G_N$ -- that restored the larger asymptotic symmetry group. While this might look strange when comparing AdS spacetimes with and without radiation, it is what we would expect from the Carrollian limit~\cite{Dappiaggi:2004kv, Ciambelli:2018wre, Donnay:2022aba,Donnay:2022wvx,deGioia:2023cbd}. 

\paragraph{An Intrinsically Carrollian Stress Tensor}
In section~\ref{sec:flat} we saw that while the flat limit of the conservation law for the stress tensor $\nabla^a T_{ab}=0$ implied the constraint equations at null infinity~\eqref{constraints}, the flat limit of the stress tensor itself was ill defined. This was first notices in \cite{Ciambelli:2018wre} in the context of the fluid/gravity correspondence. Not only were the components of the boundary stress tensor singular in the $\Lambda\rightarrow 0$ limit but, more importantly, different orders in $\Lambda$ mixed in the conservation law. While this didn't interfere with our ability to extract the celestial currents, it does illustrate a complication with defining a conserved, intrinsically Carrollian, stress tensor. Indeed the Carrollian stress tensor proposed in~\cite{Donnay:2022aba} is sourced by fluxes that are themselves functions of the radiation. It would be of great interest to identify a Carrollian stress tensor at null infinity, whose divergence reproduces the constraint equations. While our flat limit doesn't illuminate this, because there is no Levi-Civita theorem on a null hypersurface,\footnote{See the Appendix of \cite{Ciambelli:2023xqk} for a recent take on Carrollian connections.} one has more freedom in the choice of connection at null infinity, making it a worthwhile pursuit.

\paragraph{A Rigorous Analysis of AdS Radiation}
In this manuscript, we exported to AdS the characterization of radiation performed rigorously for dS and flat spaces in \cite{Fernandez-Alvarez:2019kdd, Fernandez-Alvarez:2020hsv, Fernandez-Alvarez:2021yog, Fernandez-Alvarez:2021zmp, Fernandez-Alvarez:2021uvz, Senovilla:2022pym}. We have not investigated the physical significance of the super-momentum vector in AdS, but rather, we simply used it as a diagnosis of the presence of radiation. The construction of this vector in dS and flat replied on the fact that the normal to the boundary is timelike or null, respectively. Here, the normal is spacelike, and the relevant Hamiltonian problem is the radial problem, which ultimately connects with ideas steaming from bulk reconstruction in AdS/CFT. We plan to investigate further the physical significance of the super-momentum vector, and to properly formalize the local notion of radiation of a cut of the boundary. We expect this to be deeply intertwined with the so-called filling-in problem, that is, the reconstruction of bulk geometries from boundary data. Indeed, as discussed in \cite{BernardideFreitas:2014eoi, Gath:2015nxa} and further in \cite{Ciambelli:2018wre}, we can determine the Petrov class of the bulk solution by suitably tuning the boundary Cotton tensor. A relationship between this filling-in problem and the bulk Bel-Robinson tensor has yet to be done. It will serve as a link between the bulk Petrov properties and radiation (see \cite{Trautman:1958zdi}), and its consequences on the boundary field theory. In this regard, we plan to clarify the connection with the mathematical construction of \cite{Holzegel:2015swa}, where similar conditions to those found here are explored. Another tackling angle to understand radiation  has been explored for dS in \cite{Bonga:2023eml}. It would be rewarding to apply this to AdS and compare with our predictions.

\section*{Acknowledgements}
We thank Alex Buchel, Laurent Freidel, Yangrui Hu, Justin Kulp, Luis Lehner, Rob Leigh, Rob Myers, Rodrigo Olea, Tassos Petkou, Marios Petropoulos, and Sasha Zhiboedov for valuable discussions. We thank Adam Ball, Raphael Bousso, Miguel Campiglia, Romain Ruzziconi, Jos\'e Senovilla, Jacques Smulevici, and C\'eline Zwikel for discussions and comments on the paper. This work was supported by the Celestial Holography Initiative at the Perimeter Institute for Theoretical Physics and the Simons Collaboration on Celestial Holography. Research at the Perimeter Institute is supported by the Government of Canada through the Department of Innovation, Science and Industry and by the Province of Ontario through the Ministry of Colleges and Universities.

\appendix

\section{Testing Our Radiative Criterion on Exact Solutions}\label{appA}

By construction, whenever either the Cotton or the stress tensor vanishes, eq. \eqref{sp} is automatically zero, and the spacetime under discussion is non-radiative. Our criterion becomes interesting for solutions of Einstein equations possessing a non-vanishing Cotton tensor. In this Appendix, we contrast two such spaces: the  Taub NUT and Robinson-Trautman spaces. While the former describes a rotating yet stationary black hole, the latter  describes an accelerating black hole. Therefore, the latter is expected to be radiative while the former not, even in AdS. This is confirmed by our radiative condition.

\subsection*{Taub NUT}

We follow here the notations of \cite{Kalamakis:2020aaj}, see also \cite{Ciambelli:2020qny}. The boundary of Taub NUT spaces in AdS is given by the line element
\beq
\rd s^2=g_{\mu\nu}\rd x^\mu \rd x^\nu=-(\rd t+2n (1-cos\theta)\rd \phi)^2+L^2 \rd \Omega^2,
\eeq
where $n$ is the NUT charge, $L$ is the AdS$_4$ radius and $\rd \Omega^2$ is the sphere line element.

Introducing the unit timelike vector $u^\mu=\delta^\mu_t$, the Cotton tensor is given by
\beq
C_{\mu\nu}=\frac{n}{L^4}\left(1+\frac{4n^2}{L^2}\right)(3 u_\mu u_\nu+g_{\mu\nu}).
\eeq
Similarly, the stress tensor is
\beq
T_{\mu\nu}=\frac{M}{8\pi G L^2}(3 u_\mu u_\nu+g_{\mu\nu}).
\eeq
It is therefore explicit that they are proportional, and thus eq. \eqref{sp} is identically zero, confirming that Taub NUT spaces are not radiative.

\subsection*{Robinson Trautman}

Using the conventions established in \cite{Ciambelli:2017wou}, the boundary metric of a Robinson-Trautman spacetime in AdS is
\beq
\rd s^2=g_{\mu\nu}\rd x^\mu \rd x^\nu=-\rd t+2\frac{\rd \zeta\rd \bzeta}{k^2 P^2},
\eeq
where $P$ is a function of $x^\mu=(t,\zeta,\bzeta)$, and $k=1/L$ is the inverse of the AdS radius. 

Introducing the Gaussian curvature $K=2P^2\pa_\bzeta\pa_\zeta \ln P$, the Cotton tensor is given by 
\beq
C_{\mu\nu}=i\begin{pmatrix}
0 & -\frac{k^2}{2}\pa_\zeta K & \frac{k^2}{2}\pa_\bzeta K\\
-\frac{k^2}{2}\pa_\zeta K & -\pa_t\left(\frac{\pa^2_\zeta P}{P}\right) &0\\
\frac{k^2}{2}\pa_\bzeta K & 0 &\pa_t\left(\frac{\pa^2_\bzeta P}{P}\right) 
\end{pmatrix},
\eeq
while the stress tensor reads
\beq
T_{\mu\nu}=\frac1{16\pi G}\begin{pmatrix}
4k^2 M & \pa_\zeta K & \pa_\bzeta K\\
\pa_\zeta K & \frac2{k^2}\pa_t\left(\frac{\pa^2_\zeta P}{P}\right) & \frac{2M}{P^2}\\
\pa_\bzeta K & \frac{2M}{P^2} &\frac{2}{k^2}\pa_t\left(\frac{\pa^2_\bzeta P}{P}\right) 
\end{pmatrix}.
\eeq
Here, $M$ is the black hole mass, which is time dependent as the latter is accelerated. The bulk metric is on-shell provided this stress tensor is covariantly conserved, which enforces the Robinson-Trautman equation
\beq
2P^2 \pa_\bzeta\pa_\zeta K+12M \pa_t\ln P=4\pa_t M,
\eeq
notice that this equation is independent of $L$ and thus holds for any value of the cosmological constant.

We then evaluate
\beqn
P^t&=&
2k^2P^4 \pa_t\left(\frac{\pa^2_\zeta P}{P}\right) \pa_t\left(\frac{\pa^2_\bzeta P}{P}\right)-\frac{k^4}{2}P^2\pa_\zeta K\pa_\bzeta K  \\
P^\zeta&=&-k^4 P^4\pa_t\left(\frac{\pa^2_\bzeta P}{P}\right)\pa_\zeta K+\frac{3k^6}{2} P^2 M\pa_\bzeta K\\
P^\bzeta&=&-k^4 P^4\pa_t\left(\frac{\pa^2_\zeta P}{P}\right)\pa_\bzeta K+\frac{3k^6}{2} P^2 M\pa_\zeta K.
\eeqn
We thus see that our criterion indicates there is radiation at the boundary of Robinson-Trautman, as expected. We also see that, if $\pa_t P=0$ and $\pa_t M=0$, then the Robinson-Trautman equation gives that $K$ must be the sum of a holomorphic and anti-holomorphic function. Furthermore, the condition of absence of radiation is simply given by $K$ being constant which, when $\pa_t P=0$, can always be achieved via a diffeomorphism of the $2$-sphere.

\bibliographystyle{uiuchept}
\bibliography{references.bib}

\end{document}